\documentclass[%
reprint,
superscriptaddress,
 amsmath,amssymb,
prb,
]{revtex4-2}

\usepackage[utf8]{inputenc}
\usepackage[hidelinks]{hyperref}
\usepackage[english]{babel}
\usepackage{blindtext}
\usepackage{libertine}
\usepackage{libertinust1math}
\usepackage{graphicx}
\usepackage{dcolumn}
\usepackage{bm}
\usepackage{upgreek}
\usepackage{braket}

\def\equationautorefname{Eq.}

\def\equationautorefname~#1\null{Eq.~(#1)\null}

\begin{document}
\title{Ground State Microwave-Stimulated Raman Transitions and Adiabatic Spin Transfer in the $^{15}\textrm{Nitrogen-Vacancy}$ Center}

\author{Florian B\"ohm}
\email[Author e-mail address: ]{Florian.Boehm@physik.hu-berlin.de}
\affiliation{Institut f\"ur Physik, Humboldt-Universit\"at zu Berlin, Newtonstr. 15, 12489 Berlin, Germany}
\affiliation{IRIS Adlershof, Humboldt-Universit\"at zu Berlin, Zum Großen Windkanal 6, 12489 Berlin, Germany}

\author{Niko Nikolay}
\affiliation{Institut f\"ur Physik, Humboldt-Universit\"at zu Berlin, Newtonstr. 15, 12489 Berlin, Germany}
\affiliation{IRIS Adlershof, Humboldt-Universit\"at zu Berlin, Zum Großen Windkanal 6, 12489 Berlin, Germany}
 
\author{Sascha Neinert}
\affiliation{Institut f\"ur Physik, Humboldt-Universit\"at zu Berlin, Newtonstr. 15, 12489 Berlin, Germany}
\affiliation{IRIS Adlershof, Humboldt-Universit\"at zu Berlin, Zum Großen Windkanal 6, 12489 Berlin, Germany}

\author{Christoph E. Nebel}
\affiliation{Nanomaterials Research Institute, Kanazawa University, Kanazawa, Ishikawa 920-1192, Japan}

\author{Oliver Benson}
\affiliation{Institut f\"ur Physik, Humboldt-Universit\"at zu Berlin, Newtonstr. 15, 12489 Berlin, Germany}
\affiliation{IRIS Adlershof, Humboldt-Universit\"at zu Berlin, Zum Großen Windkanal 6, 12489 Berlin, Germany}

\date{\today}


\begin{abstract}
Microwave pulse sequences are the basis of coherent manipulation of the electronic spin ground state in  nitrogen-vacancy (NV) centers.
In this work we demonstrate stimulated Raman transitions (SRT) and stimulated Raman adiabatic passage (STIRAP), two ways to drive the dipole-forbidden transition between two spin sublevels in the electronic triplet ground state of the NV center.
This is achieved by a multitone Raman microwave pulse which simultaneously drives two detuned transitions via a virtual level for SRT or via two adiabatic and partially overlapping resonant microwave pulses for STIRAP.
We lay the theoretical framework of SRT and STIRAP dynamics and verify experimentally the theoretical predictions of population inversion by observing the dipole-forbidden transition in the ground state of a single NV center. 
A comparison of the two schemes showed a better robustness and success of the spin swap for STIRAP as compared to SRT.
\end{abstract}

\maketitle

\section{\label{sec:intro}Introduction}

Recent advances in coherent control techniques of individual electron spins in solid-state quantum systems have lead to a broad spectrum of possible quantum technological applications \cite{Awschalom2013}.  
Preparation and detection of quantum states mainly depends on the ability to reliably manipulate quantum systems effectively and coherently \cite{Shapiro2000}.
The simplest form of coherent control would be a two-level system (qubit) exposed to a weak resonant driving field, which leads to coherent Rabi oscillations between the two states.
Systems with multiple states can either be effectively reduced to two-level systems by resonant control of individual transitions, or multiple states can be involved simultaneously by resonant or non-resonant control fields for extended coherent control of the system \cite{sola2018quantum}.
Moreover, instead of directly using the bare electronic states as a qubit, quantum information can be encoded in dressed states, which are generated by continuous driving of the quantum system \cite{Timoney2011}. This can be desirable as dressed states offer coherence protection \cite{Xu2012,Golter2014a}, for example from bath-induced magnetic noise.

The nitrogen-vacancy (NV) defect center in diamond \cite{Gruber1997a,Rondin2014} can be utilized as a model solid-state spin system which features multiple ground and excited states \cite{Gruber1997a,Rondin2014}. The NV center forms a single localized quantum system, an electronic qutrit ground and excited state which can be coherently controlled at room temperature by microwave or laser pulses \cite{jelezko2004observation}. In addition, the state of the NV center can be optically prepared and read out in a convenient way, which makes it a very interesting and broadly studied solid state spin system for a variety of quantum experiments such as quantum sensing \cite{Maze2008, Balasubramanian2008, Taylor2008, rong2018searching, Rembold2020, fujiwara2020real}, quantum information processing \cite{Wrachtrup2006, Cai2013, Dolde2013} and quantum computing \cite{Neumann2008, Zu2014}. 
Additionally, the nuclear spins associated to the NV center and its surrounding can be controlled as well. The latter often requires control of the employed isotope of the employed NV center \cite{Ohno2012}.

This broad range of possible applications the NV center offers stimulates a great interest in developing further coherent control schemes or adapting already known control schemes to the NV center, for example in the pursuit of enhancing magnetic field sensitivity \cite{Mamin2014,Fang2013,Barry2020}.

Both optical and microwave fields can be utilized for coherent control. Here, we concentrate on microwave fields, since they allow for better control of field amplitudes and phases, and avoid spontaneous-emission decoherence \cite{Ospelkaus2011}.
Additionally, microwave antenna structures can be easily integrated in diamond optical chips. Finally, applying microwaves for quantum control of superconducting qubits has been established \cite{Scarlino2019} at ultra-low temperatures, which may render advantageous for application with other diamond defect centers, such as the silicon-vacancy center \cite{Pingault2017}.   
To gain more insight into the simultaneous control possibilities which the NV center offers we look into the application of microwave control schemes involving all three spin ground states of the NV center, the microwave stimulated Raman transition (SRT) and the stimulated adiabatic Raman passage (STIRAP).
We adopt and establish the theoretical framework for SRT and STIRAP control in the ground state of the NV center and study these dynamics experimentally on a single NV center in $^{15}$N delta-doped diamond.
The STIRAP process is particularly interesting as it offers robustness against control pulse frequency and duration errors \cite{Shore2017} and furthermore an 
interrupted STIRAP process can be used to controllably dress and undress the NV states \cite{Timoney2011}.
In contrast to previous works on optical Raman transitions in the NV center between two ground state spin sublevels with an excited state \cite{Golter2014,Tian2019} or microwave Raman transitions between synthetic Floquet levels \cite{Shu2018}, this work exploits the full spin-1 nature of the NV center, demonstrating the full control of the dipole-forbidden transition between two spin sublevels via detuned or adiabatically overlapping microwave pulses.

The methods introduced in this work expand the already rich repertoire of control schemes for the NV center and could be an important ingredient for future quantum information, simulation or sensing schemes using the NV center.

\begin{figure*}[htbp]
  \centering
  \includegraphics[width=0.9\textwidth]{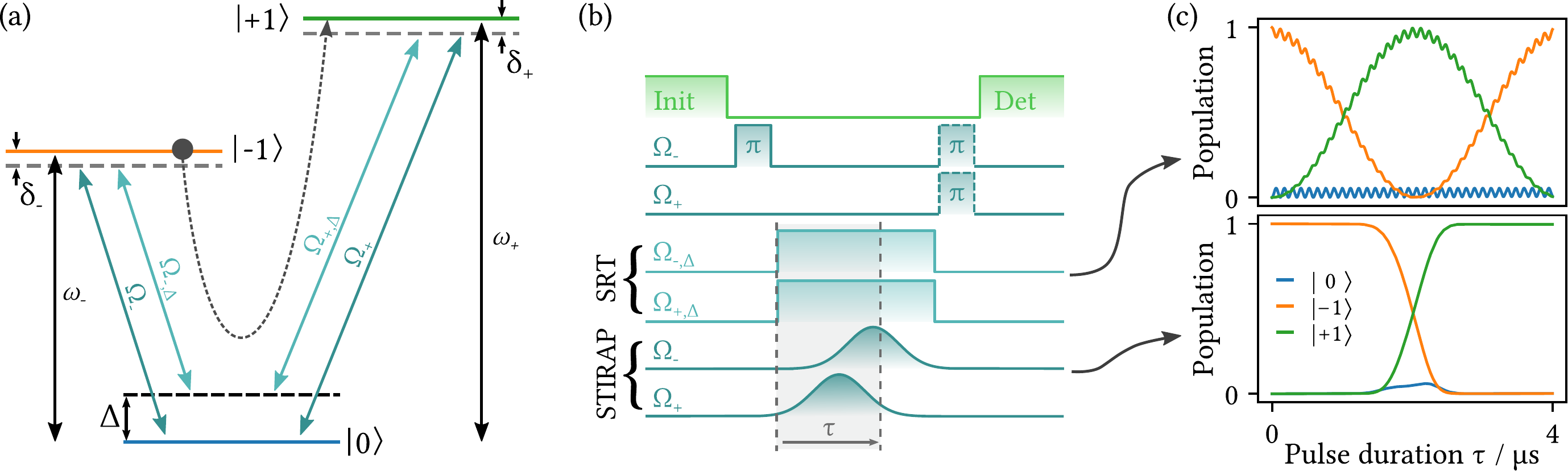}
\caption{\textit{Schematics for stimulated Raman transitions (SRT) and stimulated Raman adiabatic passage (STIRAP).}
(b) Pulse sequence for SRT and STIRAP. After polarizing the NV center into $\ket{0}$ with a green laser, the initial spin state is prepared with a resonant $\uppi_-$ pulse, swapping the population from $\ket{0}$ to $\ket{-1}$. Then the two Raman microwave pulses are applied for duration $\tau$, either detuned by $\Delta$ from the resonant transition frequency to state $\ket{0}$ for SRT or applied adiabatic and partially overlapping for STIRAP.
Afterwards either the population in state $\ket{0}$ can directly be read-out, or a $\uppi_\pm$ pulse (shown in dashed lines) resonant to one of the two transitions is applied, swapping the population between $\ket{\pm 1} \leftrightarrow \ket{0}$ to read out the population in $\ket{+1}$ or $\ket{-1}$.
(c) Example dynamic evolution of the population of the three states $\ket{0}$ and $\ket{\pm 1}$ as a function of the duration of two SRT or STIRAP pulses $\tau$ with color-coding matching that in (a).
}
\label{fig:fig1}
\end{figure*}


\section{\label{sec:theory}Theoretical Framework}
A single, negatively charged NV center's electronic spin ground state $^3A$ forms a triplet manifold, which can be initialized, manipulated and read-out conveniently at room temperature using laser and microwave fields \cite{Doherty2013}. 
The relevant Hamiltonian for the ground state of the NV center including the coupling to the $^{15}$N nucleus ($I=1/2$) can be written as:
\begin{equation}
    \mathcal{H}_{\text{NV}} = D S_z^2 + \gamma_e B_z S_z + \gamma_n B_z I_z + A S_z I_z. \label{eq:Hamiltonian_NV}
\end{equation}{}
The zero-field energy splitting between the $m_S=0$ and $m_S=\pm1$ (hereinafter referred to as $\ket{0}$ and $\ket{\pm 1}$) spin sublevels is $D=2.87\,\text{GHz}$ and the degeneracy between $\ket{\pm 1}$ can be lifted by applying an external magnetic field $B_\text{z}$ along the symmetry axis of the NV center ([111] crystal axis, we refer to this axis as $z$-axis). 
$S_z$ and $I_z$ are the z-components of the electronic and nuclear spin operator, $\gamma_e = 28.0\,$GHz$\,$T$^{-1}$ and $\gamma_n = 4.32\,$MHz$\,$T$^{-1}$ are the electronic and nuclear gyromagnetic ratios, and $A=3.03\,\text{MHz}$ is the hyperfine interaction strength for $^{15}$N.
The hyperfine coupling results in an additional energy splitting of each ground state, which is undesirable for the observation of SRT or STIRAP (see Supplemental Material at \cite{Florian2020} for more details on the influence of nuclear spin polarization). 
Fortunately a high degree of nuclear spin polarization can be achieved by working close to a level anticrossing (LAC) \cite{Jacques2009} 
or by spin polarization transfer schemes \cite{Pagliero2014}.

Disregarding the multiple nuclear spin states, the ground state of an NV center can be described by a $V$-type system as shown in \hyperref[fig:fig1]{\autoref{fig:fig1}(a)}, where the states $\ket{0} \leftrightarrow \ket{\pm 1}$ are dipole-coupled and the transition $\ket{-1} \leftrightarrow \ket{+1}$ is dipole-forbidden. 
The transition energy between the $\ket{0}$ and $\ket{-1}$ and $\ket{+1}$ state is $\omega_-$ and $\omega_+$ respectively ($\hbar = 1$) and the resonant driving fields are $\Omega_-$ and $\Omega_+$.
The primary aim of SRTs and STIRAP is to drive a direct transition between the states $\ket{-1} \leftrightarrow \ket{+1}$ without or with only little population transfer to the state $\ket{0}$ (indicated by the black dashed arrow).
For SRT this is usually achieved by applying two driving fields detuned by $\Delta$, where STIRAP uses adiabatic, partially overlapping pulses. The $\delta_\pm$ labelled detunings in \hyperref[fig:fig1]{\autoref{fig:fig1}(a)} indicate additional, unintentional detunings, which can occur due to a variety of reasons and will be discussed more in detail later.

In \hyperref[fig:fig1]{\autoref{fig:fig1}(b)}  the implemented protocol to drive and detect microwave SRT and STIRAP can be found.
First, a $\sim 4\,\text{µs}$ green ($520\,$nm) laser pulse
followed by a resonant microwave $\pi_-$ pulse ($\sim 1\,\text{µs}$)
initializes the system into the $\ket{-1}$ state.
To drive SRTs two off-resonant driving fields $\Omega_{+,\Delta}$ and $\Omega_{-,\Delta}$, each detuned by $\Delta$ from the respective resonant dipole transition $\omega_+$ and $\omega_-$  (see \hyperref[fig:fig1]{\autoref{fig:fig1}(a)}) are applied simultaneously for a varying time $\tau$. 
For STIRAP two resonant Raman control pulses with Gaussian envelope are applied in an adiabatic and partially overlapping way, where the $\ket{0}\leftrightarrow\ket{+1}$ driving field $\Omega_{+}$ is applied earlier than the $\ket{0}\leftrightarrow\ket{-1}$ driving field $\Omega_{-}$.  
The field amplitudes of the two Raman microwaves are in either case carefully adjusted to yield a nearly identical Rabi frequency $\Omega_{+,(\Delta)}  \approx  \Omega_{-,(\Delta)}$.

Subsequent to the (detuned) Raman pulses, the population in each of the three spin states can be read out by either directly reading out the NV fluorescence (the population in $\ket{0}$ is read out) or applying another state-selective resonant microwave $\pi_+$ or $\pi_-$ pulse ($\sim 1\,\text{µs}$) before reading out the fluorescence (the population in $\ket{+1}$ or $\ket{-1}$ is read out, respectively).

An example of the theoretic dynamic evolution of the three ground-states, induced by two detuned SRT microwave pulses or two adiabatic STIRAP pulses is shown in \hyperref[fig:fig1]{\autoref{fig:fig1}(c)}.
The dynamics shown here are calculated by unitary time-evolution of the incident state vector
which starts, as indicated in \hyperref[fig:fig1]{\autoref{fig:fig1}(a)}, with a polarized nuclear and electron spin in the state $\ket{-1}$. For SRTs we subsequently apply the multitone Raman MW driving field:
\begin{equation}
\begin{split}
    V(t)_{\text{SRT}} &=  B_{x,+} \sin\left( (\omega_{+} - \Delta) t \right) + B_{x,-} \sin\left( (\omega_{-} - \Delta) t \right)
       \label{eq:MW_driving} 
\end{split}
\end{equation}{}
where $B_{x,+}$ and $B_{x,-}$ is the magnetic field amplitude of the applied MW field orthogonal to the NV symmetry axis.
For STIRAP the driving field intensities are additionally modulated  by
\begin{equation}
\begin{split}
    f(t)_\pm = e^{ -(t-\mu_\pm)^2/2\sigma^2},
       \label{eq:gaussian} 
\end{split}
\end{equation}{}
which describes a Gaussian distribution with standard deviation $\sigma$ and maximum value at time $\mu_+$ and $\mu_-$ for $\omega_{+}$ and $\omega_{-}$ respectively.
Hence, the STIRAP driving field can be described by:
\begin{equation}
\begin{split}
    V(t)_{\text{STIRAP}} = B_{x,+} &\sin\left( \omega_{+} t \right) e^{ -(t-\mu_+)^2/2\sigma^2} +
    \\ B_{x,-} &\sin\left( \omega_{-} t \right) e^{ -(t-\mu_-)^2/2\sigma^2}
       \label{eq:STIRAPfield} 
\end{split}
\end{equation}{}

With the ground state Hamiltonian of the NV center $\mathcal{H}_{NV}$ (\autoref{eq:Hamiltonian_NV}) and the driving field $V(t)$ (\autoref{eq:MW_driving} or \autoref{eq:STIRAPfield}, respectively), we can write the system Hamiltonian in the lab frame as:
\begin{equation}
\begin{split}
    \mathcal{H}_{\text{NV}}^{\text{Sys}} = \mathcal{H}_{NV} &+ \left(\gamma_e S_x + \gamma_n I_x \right)  V(t)
   \label{eq:Hamiltonian_SRT}
\end{split}
\end{equation}{}
From this system Hamiltonian the evolution of the electron spin population in the three sublevels $\ket{0}$, $\ket{-1}$ and $\ket{+1}$ is computed using the Lindblad master equation solver of the {\tt QuTiP} library \cite{Johansson2013}.
In this  a dissipation process can be added by a collapse operator.
Additional collapse operators can be added to the solver to simulate decoherence.

\section{\label{sec:experiment}Experimental Implementation}
The experiments on SRT and STIRAP are carried out by observing the  spin-dependent fluorescence of a single NV center at room temperature using a home-built confocal microscope. The NV center is hosted in a [111] CVD-grown $^{15}\text{N}$ delta-doped diamond substrate and can be optically excited with a sub-nanosecond modulated diode laser at $520\,\text{nm}$ through a high numerical aperture ($\text{NA}= 1.35$) oil immersion objective. 
The $^{15}$N nuclear spin is polarized into the $m_I = +1/2$ state by a static magnetic field $B_{z} \approx 381\,\text{G}$ (close to the  excited  state  level anticrossing), applied and precisely aligned along the $z$-axis of the NV center via a permanent magnet. 
A strong hyperpolarization in $m_I = +1/2$ and good alignment of the magnetic field is verified by a vanishing electron spin resonance line of the nuclear spin state $m_I = -1/2$ (see Supplemental Material at \cite{Florian2020} for more information on nuclear spin polarization).

To drive transitions between the ground state sub-levels, the two resonant and two off-resonant microwave fields for SRT are generated by analog microwave signal generators and chopped by fast MW switches before being combined. The STIRAP sequence is directly synthesized by an arbitrary waveform generator. All MW fields are amplified and sent to a $50\,\text{µm}$ thick copper wire, which is brought in close proximity to the NV center (see Supplemental Material at \cite{Florian2020} for more details on the experimental setup). To mitigate random and systematic fluctuations in detected photons and slow drifts, the collection of data points is randomized and each measurement sequence is additionally corrected for by a reference sequence (see Supplemental Material at \cite{Florian2020} for more details on the measurement sequence).



\subsection{\label{sec:SRTs}Stimulated Raman Transitions (SRT)}
\begin{figure}[thbp]
  \centering
  \includegraphics[width=1\columnwidth]{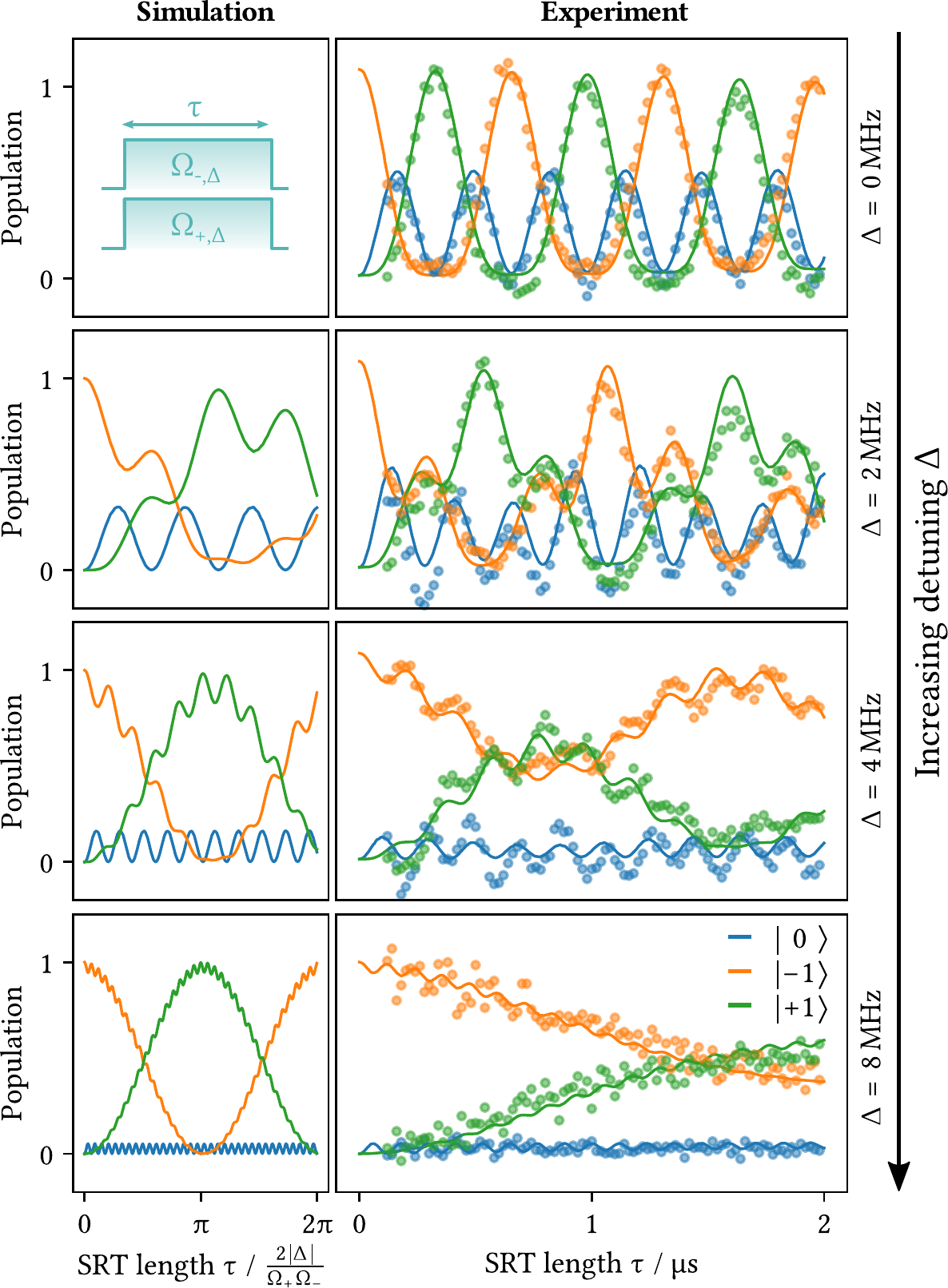}
\caption{\textit{Simulation (left) and experimental results (right) for stimulated Raman transitions (SRT).}
From top to bottom the detuning of the drive fields is increased from $\Delta=0\,\text{MHz}$ to $\Delta=8\,\text{MHz}$ as indicated on the right.
The left column shows the simulated ideal evolution of the three states in units of effective Rabi frequency $\Omega_{SRT} = \Omega_+ \Omega_- /(2 |\Delta| )$ up to $\Omega_{SRT} = 2\uppi$ (with $\Omega_{\pm,\Delta} = 2\,\text{MHz}$).
The data points in the right column show the experimentally measured populations of the three ground states $\ket{0}$, $\ket{-1}$ and $\ket{+1}$ (blue, orange, and green, respectively) as a function of the Raman pulse length $\tau$, and the data were used for simultaneous fitting of simulations (solid lines).
In the experiment, the Rabi frequency of the two Raman pulses is adjusted to $\Omega_{\pm,\Delta}\approx 2\,\text{MHz}$.
}
\label{fig:figSRT}
\end{figure}

In \hyperref[fig:figSRT]{\autoref{fig:figSRT}} we show a combination of results from simulations and experiments of SRT in the NV center.
As outlined in \autoref{sec:theory}, two detuned control fields $\Omega_{\pm,\Delta}$ are switched on simultaneously for length $\tau$ to drive SRTs, as shown in the upper left corner of \hyperref[fig:figSRT]{\autoref{fig:figSRT}}.

\begin{figure*}[thbp]
  \centering
  \includegraphics[width=1
  \textwidth]{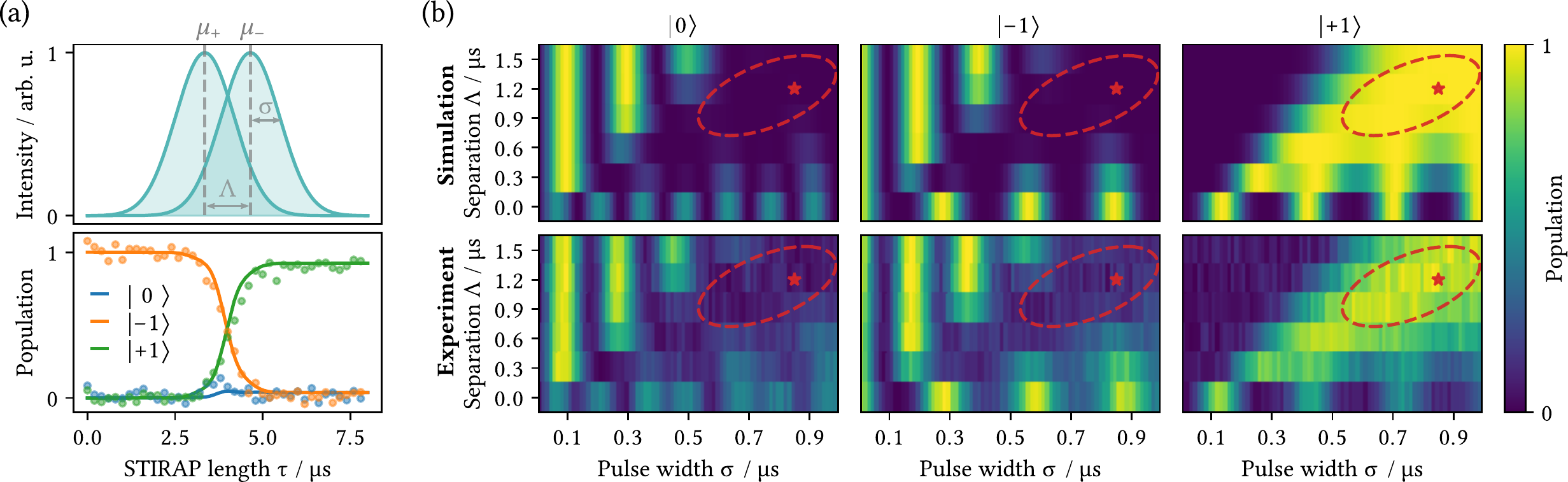}
\caption{\textit{Stimulated Raman Adiabatic Passage (STIRAP) in the ground state of the NV center.}
(a) 
The upper plot shows the normalized Gaussian profile of the two resonant STIRAP control fields $\Omega_{-}$ and $\Omega_{+}$. The Gaussian envelopes have equal width $\sigma$ and maximum value at time $\mu_-$ and $\mu_+$ respectively.
The lower plot shows experimental (data points) and simulated (solid lines) values for the population evolution in the NVs ground states $\ket{0}$, $\ket{-1}$ and $\ket{+1}$ subject to above STIRAP fields with maximum Rabi frequency
$\Omega_{\pm}\approx2\,\text{MHz}$, pulse separation $\Lambda = \mu_- - \mu_+ =  1.2\,$\textmu s and pulse width $\sigma =0.85\,$\textmu s.
(b)
The heat maps show the simulated (top row) and experimentally detected (bottom row) final population in the three ground states $\ket{0}$ (left), $\ket{-1}$ (middle) and $\ket{+1}$ (right) after STIRAP control, depending on separation $\Lambda$ and width $\sigma$ of the Gaussian control pulses. The asterisk marks the values from (a) and the red dashed area marks the broad range in which almost complete STIRAP transfer to $\ket{+1}$ occurs in simulation as well as in the experiment, thus visualizing the robustness of STIRAP against pulse width errors.
}
\label{fig:STIRAP1}
\end{figure*}

The left column of \hyperref[fig:figSRT]{\autoref{fig:figSRT}} shows the simulated ideal evolution for the population in the three states $\ket{0}$, $\ket{-1}$ and $\ket{+1}$ depending on SRT pulse length without any decoherence mechanisms and unintentional detuning $\delta_\pm = 0\,$MHz.
These calculations show that the population into the $\ket{0}$ state and the effective Raman inversion frequency $\Omega_{SRT}$ are decreasing for increasing detuning $\Delta$ (note that the detuning $\Delta$ changes the scaling of the x-axis in the plots).
In the limit of large detunings $\Delta$ the system can then be in principle reduced to a two-level spin system with an effective Rabi frequency for the spin flip Raman transition of $\Omega_{SRT} = \Omega_+ \Omega_- /(2 |\Delta| )$ \cite{Sweeney2011}.



The right column of \hyperref[fig:figSRT]{\autoref{fig:figSRT}} shows experimental results (data points) and their respective theoretical evolution (solid lines) obtained by simulations.
In the experiment, the amplitude of the two detuned Raman pulses is carefully adjusted to drive each of the two transitions with a Rabi frequency $\Omega_{\pm}\approx2\,\text{MHz}$, at the respective resonant transition  $\omega_{+}$ and $\omega_{-}$.
In contrast to the simulations in the left column of  \hyperref[fig:figSRT]{\autoref{fig:figSRT}}, the fits to the data are made considering all three spin states and including decoherence and errors in the amplitude of the two detuned Raman pulses (Rabi frequencies, $\Omega_{\pm,\Delta}$) as well as errors in detuning ($\delta_\pm$).
In the fitting process, simulations of the spin dynamics with varying unintentional detuning $\delta_\pm$ and decoherence are performed and the obtained data is fitted to the experimental data points for all three spin states simultaneous, where the quality of the fit is evaluated and the parameters are accordingly adapted until the best values are found.

For $\Delta = 0$ three-state Rabi oscillations can be seen where the population in $\ket{0}$ reaches about 0.5, while at higher detunings the population in $\ket{0}$ is drastically reduced since the system is coherently driven from $\ket{-1}$ to $\ket{+1}$ and back. 
At higher detuning $\Delta$, a reduction of coherence between $\ket{-1}$ and $\ket{+1}$ can be observed, which among other typical decoherence channels \cite{ajisaka2016decoherence} can arise from errors in $\Omega_{\pm,\Delta}$ and $\delta_\pm$ (which in turn can be caused by magnetic field drift, see Supplemental Material at \cite{Florian2020}), complicating the experimental implementation of SRTs at higher magnetic fields.

\subsection{\label{sec:STIRAP}Stimulated Raman Adiabatic Passage (STIRAP)}
For STIRAP control \cite{Gaubatz1990, Bergmann1998, Shore2017}, the two Raman control pulses are applied adiabatically and partially overlapping to produce a complete population transfer between two quantum states without or with only barely populating a (short-lived) intermediate state.
STIRAP has the advantage of being 'robust' as it relies on adiabatic change of the rotating-wave approximation Hamiltonian, which makes it less sensitive to variations e.g. of the control pulse (detuning, duration, shape, area) \cite{Shore2017} or, as in our case, changes of the magnetic field strength, which in turn can be caused by mechanical drift of the sample with respect to the employed permanent magnet.

In the upper part of \hyperref[fig:STIRAP1]{\autoref{fig:STIRAP1}(a)} the Gaussian envelope of the two STIRAP control fields used in the experiment and described in \autoref{eq:gaussian} is illustrated. 
In STIRAP the two pulses are applied in a seemingly \textit{counter-intuitive} ordering, which in our case means the $\ket{0}\leftrightarrow\ket{+1}$ driving field $\Omega_{+}$ is applied first with peak at $\mu_+$ (even though there is no initial population in state $\ket{0}$ or $\ket{+1}$) and 
the second driving field $\Omega_{-}$, resonant to the transition $\ket{0}\leftrightarrow\ket{-1}$ and with peak at $\mu_-$ is applied subsequently.
The separation between the peaks of the two Gaussian envelopes is $\Lambda = \mu_- - \mu_+$ and the two Gaussian pulses are set to the same width and peak amplitude, as well as temporal pulse area, hence integrated Rabi frequency. 
After a width of $w = 8 \sigma$ the amplitude of each microwave field is set to $0$ both in the numerical simulation as well as in the experiment.

The lower part of \hyperref[fig:STIRAP1]{\autoref{fig:STIRAP1}(a)} shows  experimental results for the dynamic evolution (data points) of the three ground states $\ket{0}$, $\ket{-1}$ and $\ket{+1}$ (blue, orange, and green, respectively) of the NV center, after applying the two previously described resonant STIRAP Gaussian shaped drive fields.
For this experiment, the STIRAP sequence was clipped (which means that the amplitude of the drive fields is suddenly set to $0$) after time $\tau$ as shown in \hyperref[fig:fig1]{\autoref{fig:fig1}(b)} to visualize the STIRAP transfer dynamics. It can be clearly seen that population transfer from $\ket{-1}$ to $\ket{+1}$ occurs with almost no population in $\ket{0}$, a signature of traditional STIRAP.
The experimental results also fit well with theoretically calculated dynamics (see \autoref{sec:theory}) including decoherence (solid line).

In STIRAP the focus lies in the robustness and success of the spin swap, therefore the six plots arranged in tabular form in \hyperref[fig:STIRAP1]{\autoref{fig:STIRAP1}(b)} show the final population in each of the three ground states, depending on the separation of the two Gaussian control pulses $\Lambda$ and the the standard deviation width $\sigma$ of the pulses.
The three columns represent the final population in $\ket{0}$,  $\ket{-1}$ and $\ket{+1}$ where the top row shows numerically simulated and the bottom row experimental data.
Here it can be seen that the population shows plateaus in both simulation and experimental data where a complete population transfer can occur for a wide range of pulse widths (final population in $\ket{0} \rightarrow 0$, $\ket{-1} \rightarrow 0$ and $\ket{+1} \rightarrow 1$, red dashed marked areas as guide for the eye), showing that microwave STIRAP can be a very robust method to drive the spin-forbidden $\ket{-1} \leftrightarrow \ket{+1}$ transition in the ground state of the NV center.
The heat maps in \hyperref[fig:STIRAP1]{\autoref{fig:STIRAP1}(b)} can also be used to choose a robust point to perform STIRAP experiments, in this case the asterisk in the heat maps mark the point where the dynamic STIRAP evolution in \hyperref[fig:STIRAP1]{\autoref{fig:STIRAP1}(a)} was recorded, slight changes in the microwave pulses should have no strong effects on the outcome of the sequence around this point. Furthermore, from the dynamic STIRAP evolution we can infer that almost no population is transferred into $\ket{-1}$ during STIRAP evolution in the marked area.

\subsection{\label{sec:comparison}Robustness of SRT and STIRAP}
\begin{figure}[thbp]
  \centering
  \includegraphics[width=1\columnwidth]{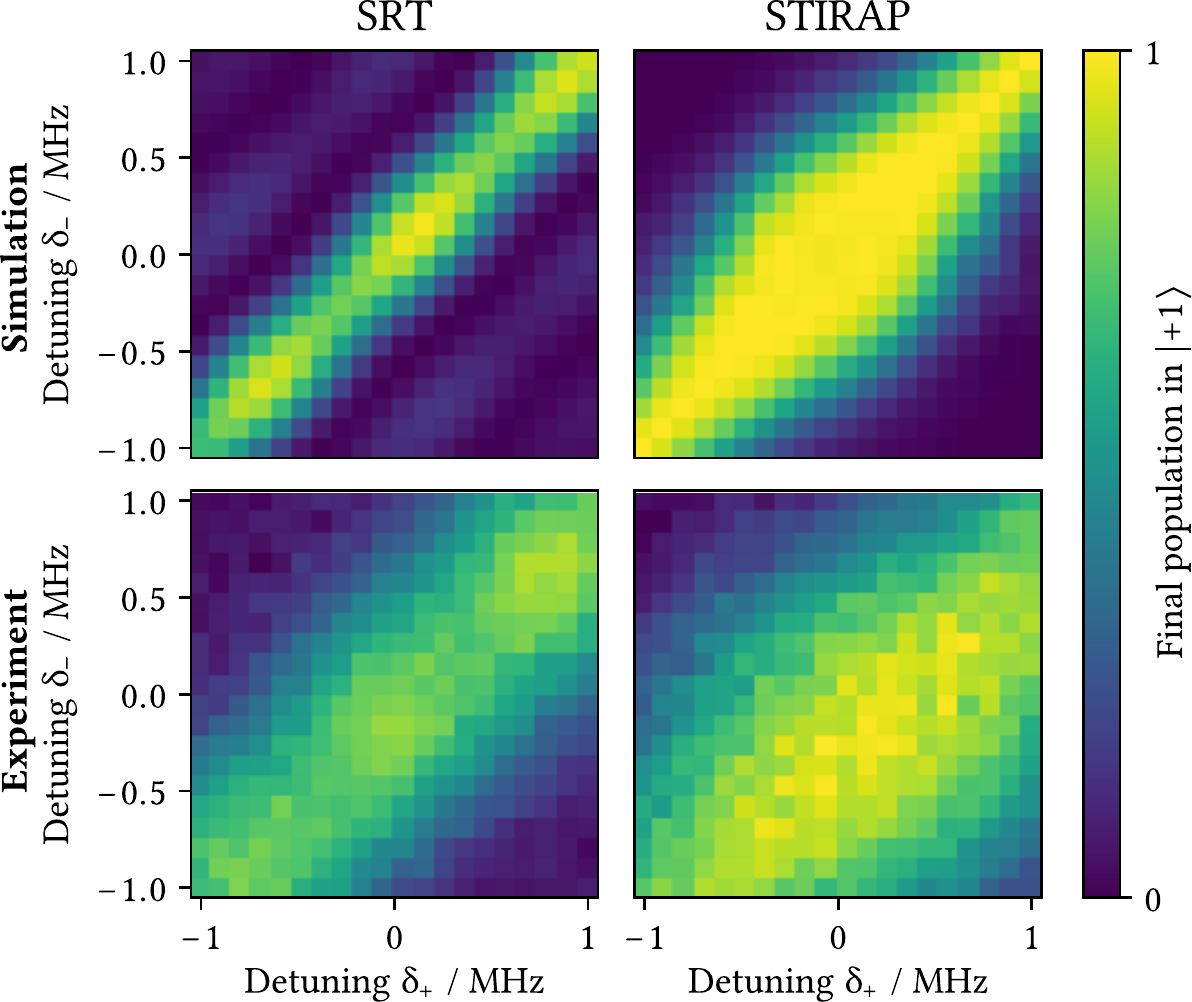}
\caption{\textit{Comparison of robustness of SRT and STIRAP against frequency errors.}
The heat maps represent the final population of the desired state $\ket{+1}$ with SRT (left column) and STIRAP (right column) control in dependence of an unintentional frequency detuning $\delta_\pm$ of the two control pulses. The top row shows simulation results and the bottom row shows experimental results. SRT were chosen with an initial detuning $\Delta = 5\,$MHz and STIRAP with no initial detuning, Gaussian RMS width $\sigma=0.85\,$\textmu s and pulse separation $\Lambda = 1.2\,$\textmu s. The simulation shows much sharper patterns than the experimental data, which is most likely due to  frequency jitter occurring during acquisition of data. Nevertheless, a strong dependence of SRT on frequency errors (which can occur due to changes in applied magnetic field, temperature or drift) and the higher robustness of STIRAP against such errors, i.e. a larger area where spin flip was successful, are apparent.
}
\label{fig:STIRAP2}
\end{figure}

STIRAP also promises to be to more robust to unintentional frequency detuning of the two control pulses ($\delta_{\pm}$, see \hyperref[fig:fig1]{\autoref{fig:fig1}(a)}), meaning an additional offset to the intentionally detuned SRT pulse frequency or the resonant STIRAP pulse frequency.
The effect of this frequency detuning on the outcome of the SRT or STIRAP sequence are presented in \hyperref[fig:STIRAP2]{\autoref{fig:STIRAP2}},
where detuning maps compare the final population in the state $\ket{+1}$ with respect to frequency detuning of the two control pulses $\delta_{\pm}$ for SRT and STIRAP.

The top row of \hyperref[fig:STIRAP2]{\autoref{fig:STIRAP2}} compares the outcome of a simulated SRT and STIRAP sequence into the desired final state $\ket{+1}$.
It is evident from these simulations that in case of SRT a small frequency error ($\delta \ll 1\,$MHz) already has a strong effect on the outcome of the sequence, whereas STIRAP shows a much higher robustness against frequency changes. 
The bottom row of \hyperref[fig:STIRAP2]{\autoref{fig:STIRAP2}} shows the corresponding experimental data. 
A systematic deviation from of the experimental data from the simulation can be observed, which is most likely due to frequency drifts during the
acquisition of the data. A 2D detuning map with $20 \times 20 = 400$ individual data points each integrated for about $0.5-1.0\,$s usually takes about $5\,$min for a single acquisition. This is then repeated about 50 times in order to obtain 2D detuning maps with low noise, as they can be found in \hyperref[fig:STIRAP2]{\autoref{fig:STIRAP2}}. However, even when regarding only the data from the first integration 
the sharp features as expected from simulation can not be found, therefore the frequency fluctuations appear to be on a timescale $<5\,$min.
However, the data still confirms that the STIRAP sequence is less sensitive to frequency changes and suffers less from decoherence.

Frequency drifts can occur internal or external to the NV center, meaning either the resonance frequency of the NV itself shifts, where one reason for this can be thermally induced lattice strains \cite{Acosta2010} or the actual magnetic field applied to the NV center changes e.g. caused by mechanical drifts (see Supplemental Material at \cite{Florian2020} for more information on magnetic field drifts).
Especially in the case of SRT we can see that drifts of the applied magnetic field (which shifts $\delta_+$ and $\delta_-$ in opposite directions) can aggravate the outcome of the sequence drastically.

\section{\label{sec:conclusion}Conclusion and outlook}
Summarizing, we show the microwave stimulated Raman transitions (SRT) and stimulated Raman adiabatic passage (STIRAP) in the triplet ground state of the NV center. This is achieved either by using a multitone microwave pulse that simultaneously drives two detuned ground state transitions for SRT or applying two partially overlapping resonant adiabatic pulses for STIRAP.
We elucidated the theoretical framework for SRT and STIRAP ground state spin dynamics and experimentally observed both SRT and STIRAP dynamics, demonstrating the ability to drive the spin-forbidden $m_S =-1$ to $m_S=+1$ transition with almost no population transfer to the $m_S=0$ level in both cases.

SRT prove to be technically more easy to realize as no adiabatic pulses are required, but suffer more from decoherence, especially for higher mean detuning and also are experimentally more challenging due to high sensitivity to frequency errors, which can for example occur from mechanical drifts. STIRAP requires adiabatic pulse shapes, which are technically more demanding to generate, but proves to be more robust to both pulse area deviations and frequency fluctuations and also suffers less from decoherence, hence rendering it simpler to realize experimentally.

The possibility to drive microwave SRT and STIRAP transitions in the ground state of the NV center and thus driving the spin-forbidden transition of the system via microwaves could open up new possibilities for quantum control schemes of the NV center, which could be applied for example in future quantum sensing schemes. Furthermore the microwave STIRAP sequence could be adapted to generate dressed states in the NV center, offering coherence protection.

\section{Acknowledgments}
We acknowledge financial support by the Federal Ministry of Education and Research of Germany in the framework of Q.Link.X (project number 16KIS0876).
The simulations are coded in {\tt PYTHON} using the {\tt QuTiP} library \cite{Johansson2013}.
%

%
\end{document}


\title{Supplementary Information\\
Ground State Microwave-Stimulated Raman Transitions and Adiabatic Spin Transfer in the $^{15}\textrm{Nitrogen-Vacancy}$ Center}


\author{Florian B\"ohm}
\email[Author e-mail address: ]{Florian.Boehm@physik.hu-berlin.de}
\affiliation{Institut f\"ur Physik, Humboldt-Universit\"at zu Berlin, Newtonstr. 15, 12489 Berlin, Germany}
\affiliation{IRIS Adlershof, Humboldt-Universit\"at zu Berlin, Zum Großen Windkanal 6, 12489 Berlin, Germany}

\author{Niko Nikolay}
\affiliation{Institut f\"ur Physik, Humboldt-Universit\"at zu Berlin, Newtonstr. 15, 12489 Berlin, Germany}
\affiliation{IRIS Adlershof, Humboldt-Universit\"at zu Berlin, Zum Großen Windkanal 6, 12489 Berlin, Germany}
 
\author{Sascha Neinert}
\affiliation{Institut f\"ur Physik, Humboldt-Universit\"at zu Berlin, Newtonstr. 15, 12489 Berlin, Germany}
\affiliation{IRIS Adlershof, Humboldt-Universit\"at zu Berlin, Zum Großen Windkanal 6, 12489 Berlin, Germany}

\author{Christoph E. Nebel}
\affiliation{Nanomaterials Research Institute, Kanazawa University, Kanazawa, Ishikawa 920-1192, Japan}

\author{Oliver Benson}
\affiliation{Institut f\"ur Physik, Humboldt-Universit\"at zu Berlin, Newtonstr. 15, 12489 Berlin, Germany}
\affiliation{IRIS Adlershof, Humboldt-Universit\"at zu Berlin, Zum Großen Windkanal 6, 12489 Berlin, Germany}

\date{\today}


\maketitle

\section{\label{sec:setup}Experimental Setup}
This chapter gives a more detailed overview of the optical, electrical and magnetic components of the experimental setup employed to investigate the SRT and STIRAP dynamics. The diamond used in the experiment is a CVD grown type II B diamond with a delta-doped layer of nitrogen-15 impurities.
\subsection{\label{subsec:Opticalsetup}Optical Setup}
A single isolated $^{15}$NV center can be found and individually addressed in a home-built confocal setup. For excitation a diode laser at $520\,$nm with digital modulation (Labs Electonics, DLNSEC0520) is directed and reflected towards the sample with a dichroic mirror. The excitation light is then focused on the sample using an immersion oil objective with high numerical aperture ($\text{NA}=1.35$, Olympus, UPLANSAPO60X). The excited fluorescence is collected using the same objective and on it's way back mainly passes the dichroic mirror ($\lambda > 600\,$nm) and a further longpass filter (RPE620LP, Omega Optical) and is then focused on a pinhole with a diameter of $50\,$µm (Thorlabs) for additional confocal filtering.  The filtered light is then sent to two fast avalanche photodetectors (Excelitas, SPCM-AQRH-14 APD) through a 50/50 nonpolarizing beamsplitter in a Hanbury-Brown Twiss configuration. The photon coincidence statistics can be evaluated to prove the mainly single photon emission ($g^2(0) < 0.5$), hence to verify the observation of fluorescence from a single emitter.

\subsection{\label{subsec:MWsetup}Microwave Setup}
The pulse sequence for SRT (see Fig. 1(b) of the main text) requires four microwave (MW) fields with different frequency and amplitude which need to be controlled individually. For this we use analog microwave signal generators (HAMEG, HM8135; Agilent, E8267; SRS, SG380; TTi, TGR6000) which can be controlled by our home-build software and are all synchronised to the same $10\,$MHz reference. The two Raman MW fields are first combined and then modulated by a fast RF switch (Mini-Circuits, ZASW-2-50DRA+), before being combined with the two other microwave fields, where each can be individually modulated by another fast RF switch (Mini-Circuits, ZASW-2-50DRA+). The combined MW field is then amplified by a MW amplifier (Mini-Circuits, ZHL-16W-43-S+) before being sent to the copper wire on top of the diamond.
In this case the whole experiment is synchronised by a digital arbitrary waveform generator (Swabian Instruments, Pulse Streamer 8/2), which generates TTL signals to control the pulsed laser, the MW switches and the photon detection electronics.

The pulse sequence for STIRAP (see Fig. 1(b) of the main text) requires four microwave fields with different frequency  and additionally the ability to arbitrary modulate the amplitude of these fields. For this we use an arbitrary waveform generator (AWG) (Keysight, M8195A) with four output channels. In this case the AWG generates the whole microwave sequence into one of the output ports and the other three outputs generate the digital signal to control the pulsed laser and gate the photon detection electronics.

\subsection{\label{subsec:magnet}Static Magnetic Field}
In the experiment the static magnetic field along the NV center symmetry axis $B_z$ is applied by a circular neodymium permanent magnet (N45, $r=25\,$mm, $d=8\,$mm). The magnet is mounted on the opposite side of the microscope objective's front lens and its XYZ position can be precisely controlled by three linear translation stages. The diamond sample and microwave antenna are mounted in between the magnet and the objective lens.
Assuming the position of the microscope objective and the magnet are fixed with respect to each other, a drift of the NV center in $z$-direction (the axis between magnet and objective lens) can be observed both by a shift of the piezo $z$-position that is required to have the NV center in focus and a shift of the NV electron spin resonance lines.
By observing these shifts, which can e.g. occur when the microwave power applied to the antenna mounted on the diamond is changed drastically, the distance dependence of the magnetic field could be estimated to be about $\frac{d B_z}{d z} \approx 0.1\,$G$\,$µm$^{-1}$.
The distance dependence of the magnetic field could also be verified to a very good degree  by FEMM simulations \cite{Meeker} using the magnets dimensions, material composition and distance to the sample.

\section{\label{sec:n15hyper}$^{15}$NV Verification and Nuclear Spin Polarization}
\begin{figure}[htbp]
  \centering
  \includegraphics[width=1\columnwidth]{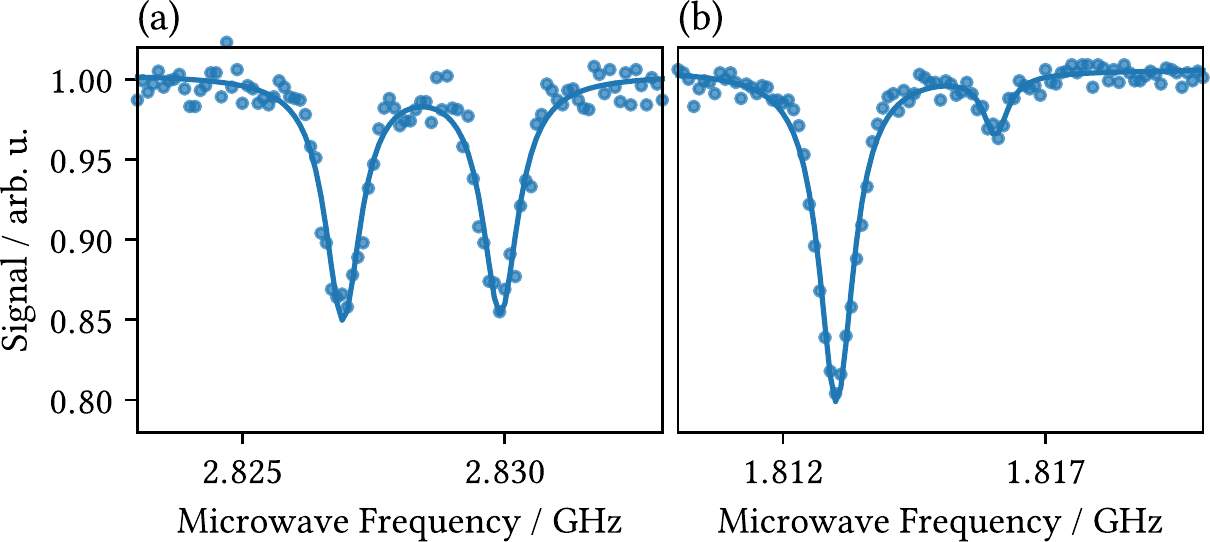}
\caption{\textit{Optically detected magnetic resonance (ODMR) spectra at low and high magnetic field magnitude.} (a) The characteristic two  electron spin resonance (ESR) lines for $^{15}$NV centers at $B_z\approx 15\,$G showing an equal population in the nuclear spin $m_I = -1/2$ (right line) and $m_I = +1/2$ (left line). (b) At higher magnetic field magnitude of $B_z\approx 380\,$G, close to the excited state level anticrossing, the $m_I = -1/2$ ESR line almost completely disappears, indicating a nuclear spin polarization in state $m_I = +1/2$. The solid lines are fits to the data points using two Lorentzian functions.
}
\label{fig:N15}
\end{figure}
The nitrogen isotope associated to the NV defect can be revealed by scanning the microwave frequency over the electron spin resonance (ESR) with relatively low microwave power (to avoid power broadening of the lines) and recording the fluorescence, recording an optically detected magnetic resonance (ODMR) spectrum. 
The energy splitting due to hyperfine interaction between nuclear and electron spin  for $^{14}$N and $^{15}$N are around $2.2\,$MHz and $3\,$MHz respectively. Furthermore, the $^{14}$N and $^{15}$N nuclei have spin $S=1$ and $S=1/2$ and therefore show three and two hyperfine lines respectively.

In \hyperref[fig:N15]{\autoref{fig:N15}(a)} we show an ODMR scan of the NV center's $m_S = 0$ to $m_S = -1$ manifold at relatively low magnetic field magnitude ($B_z\approx 15\,$G), where two resonance lines with equal contrast and spaced apart about $3\,$MHz can be found, clearly identifying the nitrogen isotope 
as $^{15}$N.

To achieve nuclear spin polarization into the $m_I = +1/2$ state a static magnetic field close to the excited-state level anticrossing (ESLAC at $B_z\approx 514$\,G) is applied and precisely aligned along the $z$-axis of the NV center. This alignment and polarization can  be verified by an ODMR scan as it can be found in \hyperref[fig:N15]{\autoref{fig:N15}(b)} by an almost vanishing ESR line for the nuclear spin state $m_I = -1/2$.

\section{\label{sec:unpolar}STIRAP and SRT with and without nuclear spin polarization}
\begin{figure}[htbp]
  \centering
  \includegraphics[width=1\columnwidth]{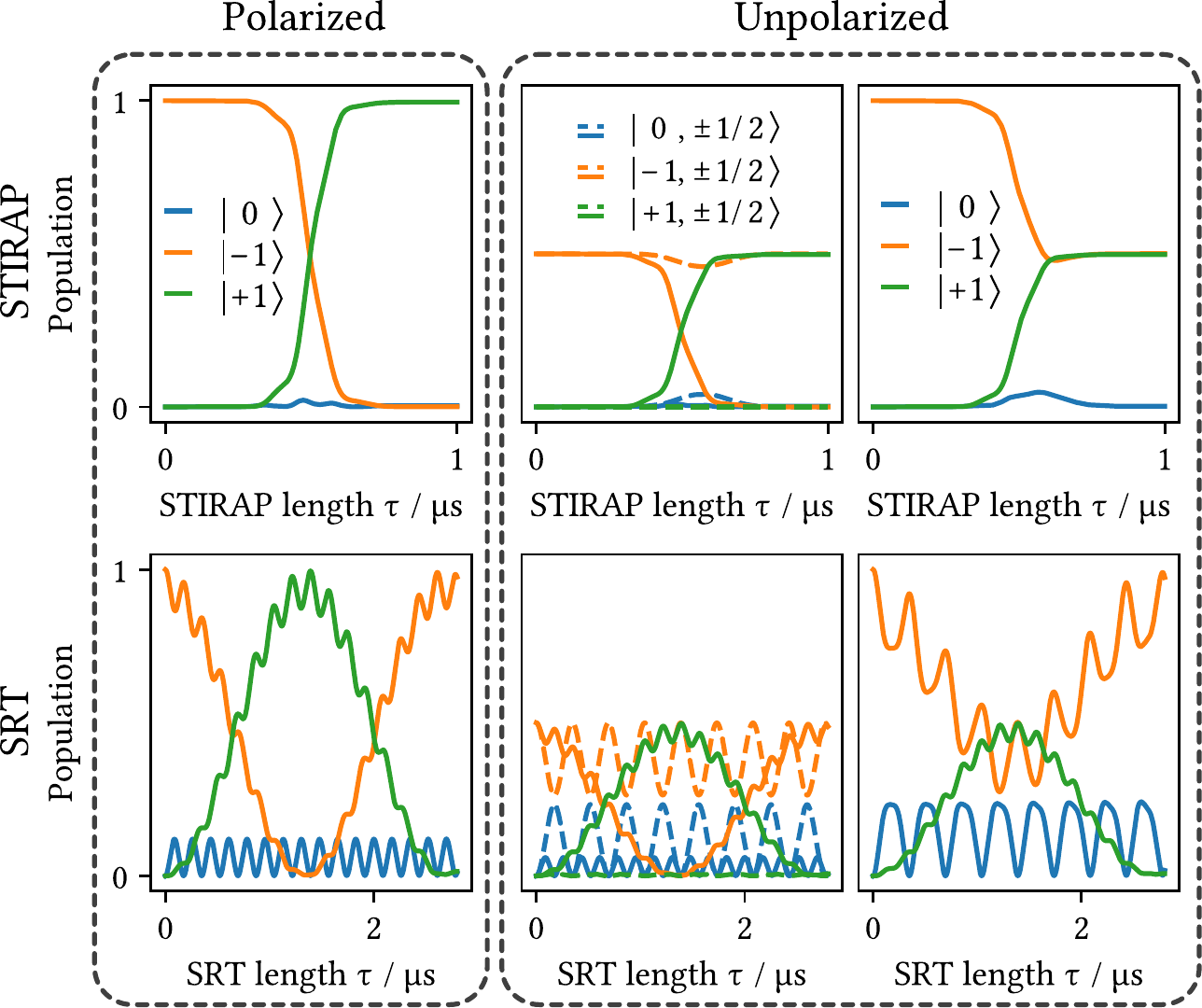}
\caption{\textit{Polarized and unpolarized nuclear spin.} The top row shows the simulated dynamics of STIRAP with $\sigma = 0.85\,$µs, $\Lambda = 1.2\,$µs, the bottom row simulated SRT with $\Delta = 5\,$MHz. The left column shows the expected dynamics with the nuclear spin  polarized in the $m_I = -1/2$ state, as it is assumed throughout this work. The middle and right column show the expected dynamics for an unpolarized nuclear spin, where the middle column shows the actual evolution of the separate nuclear spin states $m_I = \pm1/2$ and the right column shows the expected read out signal which is a sum over each of the two nuclear spin states. The dashed and solid line corresponds to the $m_I = +1/2$ and $m_I = -1/2$ spin, respectively.
}
\label{fig:unpolarized}
\end{figure}
In  \hyperref[fig:unpolarized]{\autoref{fig:unpolarized}} we show the electron spin dynamics of the NV center under SRT ($\Delta = 5\,$MHz) and STIRAP ($\sigma = 0.85\,$µs, $\Lambda = 1.2\,$µs) in the case of the nuclear spin being polarized into the $m_I = +1/2$ state (left column) and the nuclear spin being unpolarized (center and right column), hence the $m_S = -1, m_I = +1/2$ ($\ket{-1,+1/2}$) and $m_S = -1, m_I = -1/2$ ($\ket{-1,-1/2}$) state initially being equally populated. When the population of the NV ground state is optically read out it is not possible to distinguish between the two nuclear spin states $m_I=\pm 1/2$, hence the detected signal is always a combination of the two hyperfine states. Therefore, the central and right column of \hyperref[fig:unpolarized]{\autoref{fig:unpolarized}} show the expected evolution of the electronic and nuclear ground states in the case of no initial nuclear spin polarization and the expected signal to be detected, when reading out the SRT or STIRAP applied to a NV center with no initial nuclear spin polarization, respectively.

\section{\label{sec:sequence}Normalization Sequence}
\begin{figure}[thbp]
  \centering
  \includegraphics[width=1\columnwidth]{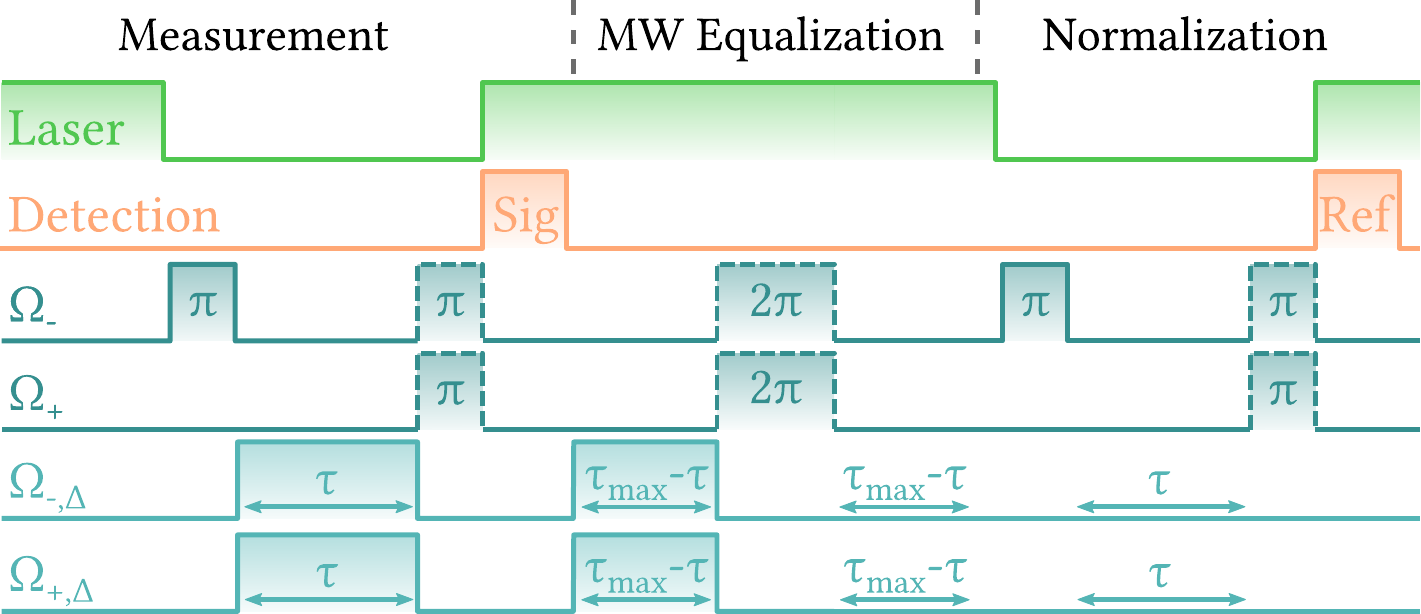}
\caption{\textit{Normalization sequence.} Sketch of the three parts of the actual measurement sequence, employed for mitigating the heating due to high-power microwaves applied to the sample, $T_1$ relaxation time and potential $\pi$-pulse errors. When the $\pi$-pulse in the measurement and normalization sequence is not applied, a $2\pi$-pulse is applied in the equalization part of the sequence and vice versa.}
\label{fig:sequencesuppl}
\end{figure}

In order to map the change in detected photons to the population of the $\ket{0}$ state, a normalization method as it is illustrated in \hyperref[fig:sequencesuppl]{\autoref{fig:sequencesuppl}} for the example of SRT, is used. 
The first part of the sequence is the measurement part, where the actual microwave sequence is running and the gated detection (labelled 'Sig') is used to detect photons arriving in a $\sim 300\,$ns window after the last microwave pulse finished and the laser is turned back on. The second part of the sequence is to mitigate heating effects due to heat contribution of the high-power (peak power $\sim 15\,$W) MW frequency applied to the antenna, which can in turn cause drifts of the sample in respect to the microscope objective and magnet. Therefore this second part of the sequence is applied to compensate for the different pulse length $\tau$ applied during the measurement part and  keep the duty cycle of the applied microwave fields during a whole sequence step similar. The third part of the sequence is similar to the first part, acquiring the reference signal during the 'Ref' window. To acquire the reference photon counts, instead of the SRT/STIRAP microwave pulses no microwave is applied for a waiting time $\tau$, mitigating for the ground state spin lifetime ($T_1$-time) of the NV center and non-perfect state transfer which could be caused by the $\pi$-pulses driving the $\ket{0} \leftrightarrow \ket{\pm1}$ transition.

\section{\label{sec:robustness}Sensitivity of SRT to fluctuations}
Here we will discuss the sensitivity/robustness of SRT to environmental fluctuations, as especially SRT suffer a lot from unintentional frequency detuning (see Fig. 4 in main text). The acquisition time for SRT measurements, detecting all three spin states, usually takes hours during which the system might be prone to mechanical  or temperature drifts. Here we focus on two major mechanisms changing the NVs magnetic-resonance spectra, temperature and position drifts and evaluate their influence on  SRT dynamics.

\begin{figure}[hb]
  \centering
  \includegraphics[width=1\columnwidth]{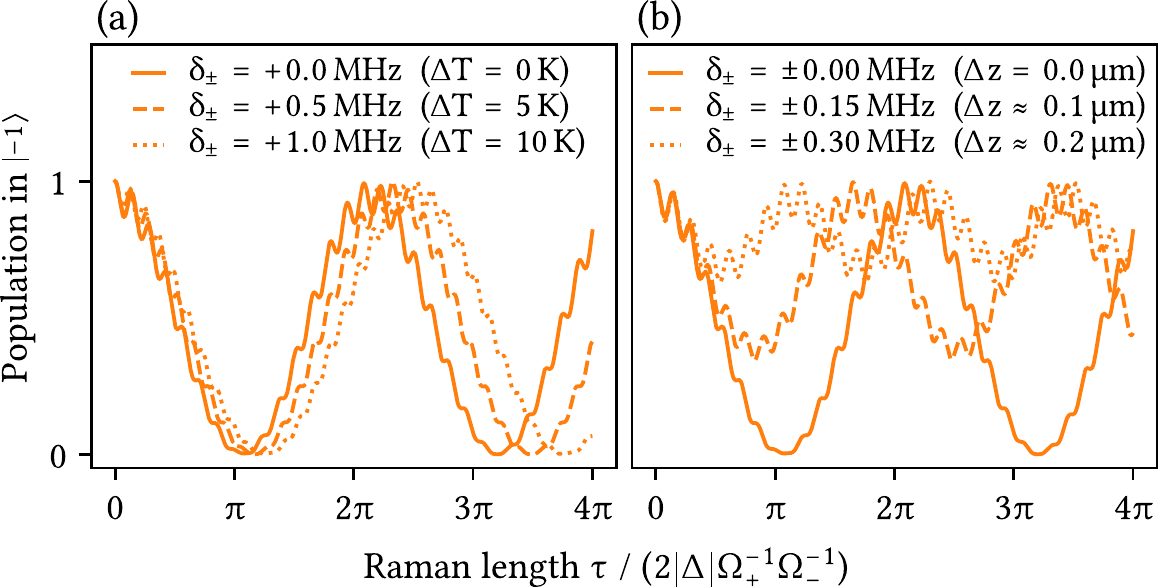}
\caption{\textit{Evaluation of environmental fluctuations.} The solid lines in (a) and (b) represent simulated SRT dynamics of state $\ket{-1}$ for $\Delta=5\,$MHz,  $\Omega_{\pm,\Delta}=2\,$MHz, $B_{z}=381\,\text{G}$ with zero unintentional detuning $\delta_\pm = 0\,$MHz.
(a) Temperature fluctuations $\Delta T$ influence the zero-field splitting of the NVs electronic ground-states and therefore a change in the mean Raman detuning $\Delta$ by $\delta_\pm$. The dashed and dotted line represent $\Delta T = 5\,K \rightarrow \delta_\pm \approx +0.5\,$MHz and $\Delta T = 10$\,$K \rightarrow \delta_\pm \approx +1.0\,$MHz respectively. 
(b) Fluctuations of the  static magnetic field $B_z$ change the two individual Raman detunings causing $\delta_+$ and $\delta_-$ to shift in opposite direction. A change in magnetic field can e.g. be caused by changes of the distance between the magnet and NV center $\Delta z$. 
The dashed and dotted line represent a shift of $\delta_\pm = \pm 0.15\,$MHz and $\delta_\pm = \pm 0.3\,$MHz respectively, corresponding to an estimated shift of the distance between  NV and magnet of $\Delta z \approx 1\,$µm and $\Delta z \approx 2\,$µm in the experiment.
}
\label{fig:SRTfluctuations}
\end{figure}

In \hyperref[fig:SRTfluctuations]{\autoref{fig:SRTfluctuations}(a)} we evaluate the sensitivity of the SRTs to temperature fluctuations.
Thermally induced lattice strains result in a temperature dependence of the NV centers zero-field transition frequency $\frac{\partial D_s}{\partial T} \approx 0.1\,\text{MHz}\,\text{K}^{-1}$ around room temperature \cite{Acosta2010}.
Temperature fluctuations $\Delta T$ therefore change the splitting between $\ket{\pm1}$ and $\ket{0}$, but both spin states $\ket{-1}$ and $\ket{+1}$ shift in the same direction with the same amount with respect to $\ket{0}$, meaning a change in $D_s$ effectively only changes the mean Raman detuning $\Delta$ but not $\delta_\pm$. 
Evaluating the SRT dynamics under temperature change (see \hyperref[fig:SRTfluctuations]{\autoref{fig:SRTfluctuations}(a)}) we find that the SRTs inversion frequency changes with temperature but still a full population inversion is predicted.

In \hyperref[fig:SRTfluctuations]{\autoref{fig:SRTfluctuations}(b)} we evaluate numerically the sensitivity of SRTs to fluctuations of the applied static magnetic field.
As the Zeeman shift of the electron spin sublevels $\ket{-1}$ and $\ket{+1}$ is directly dependent on $B_z$, the resonances of these two transitions get shifted in opposite directions when the static magnetic field along the $z$ axis changes. Hence, a change in $B_z$ results in no change of the average Raman detuning $\Delta$ but a change of $\delta_+$ and $\delta_-$ in opposite direction. 
Fluctuations of the magnetic field can for example be caused by fluctuations of an electromagnet or changes in the distance between the NV center and the source of the magnetic field.
Especially at higher magnetic fields, e.g. around the level anticrossing points the magnetic field has a high distance dependence, in our typical experimental setting around $B_z \approx 380\,$G we approximate $\frac{\partial B_z}{\partial z} \approx 0.1\,$G$\,$µm$^{-1}$ from FEMM simulations (see supplement \ref{subsec:magnet}). As it can be seen in \hyperref[fig:SRTfluctuations]{\autoref{fig:SRTfluctuations}(b)} a change in magnetic field can have drastic effects on the inversion probability, as compared to usual monochrome sensing schemes, the SRT scheme basically senses two transitions at the same time. This shows the importance to keep the experimental setup for SRT mechanically very stable or use more robust schemes like STIRAP.

%